\documentclass[twocolumn,showpacs,preprintnumbers,amsmath,amssymb]{revtex4}

\usepackage{graphicx}
\usepackage{dcolumn}
\usepackage{bm}

\begin{document}

\title{Recursive graphs with small-world scale-free properties}

\author{Francesc Comellas}
\affiliation{
Departament de Matem\`atica Aplicada IV, EPSC, Universitat
Polit\`ecnica de Catalunya\\
 Avinguda del Canal Ol\'{\i}mpic s/n, 08860
Castelldefels, Barcelona, Catalonia, Spain
}%
\homepage{http://www-mat.upc.es/~comellas}
\email{comellas@mat.upc.es}

\author{Guillaume Fertin}
\affiliation{
IRIN, Universit\'e de Nantes, 2 rue de la Houssini\`ere, 
 BP 92208, 44322 Nantes Cedex 3, France
}%
\homepage{http://www.sciences.univ-nantes.fr/info/perso/permanents/fertin/}
\email{fertin@irin.univ-nantes.fr}

\author{Andr\'e Raspaud}
\affiliation{
LaBRI, Universit\'e Bordeaux I,  351  Cours de la Lib\'eration, 
33405 Talence  Cedex, France
}%
\homepage{http://www.labri.fr/Perso/~raspaud/}
\email{raspaud@labri.fr}

\date{}
\date{Rec. 26 November 2002; revised manuscript rec. 15 December 2003; publ. 31 March 2004}

\begin{abstract}
We discuss a category of graphs, recursive clique trees,
which have small-world and scale-free properties and allow
a fine tuning of the clustering and the power-law exponent of their 
discrete degree distribution.
We determine  relevant characteristics of those graphs: 
the diameter, degree distribution,  and  clustering parameter.
The graphs have also an interesting recursive property, 
and generalize recent constructions with fixed degree 
distributions. 
\end{abstract}

\maketitle

In a complex system a large number of agents interact showing 
a cooperative behavior which does not depend only on the 
individual features of its parts, but also on its structure 
and on the different sort of relations which can be established 
between them and the environment. This global behavior allows 
the system to attain certain achievements without the presence 
of an administrative hierarchy or a central control mechanism. 
A swarm of bees, an ant colony, companies which supply a 
big city (water, electricity, telephone, etc.), even biological 
mechanisms or social relationships are all complex systems where 
global patterns emerge from the interaction of a large number 
of similar elements.  
In recent years, there has been an increase in understanding 
of such complex systems in terms of networks, modeled by graphs 
composed of vertices and edges, where vertices represent the 
basic elements and edges their interactions.
The World Wide Web (WWW), Internet, transportation systems, and many 
biological and social systems have been characterized by 
small-world scale-free networks~\cite{WaSt98,BaAl99} which 
have a strong local clustering (nodes have many mutual neighbors), 
a small diameter (maximum distance between any two nodes) and 
a distribution of the number of nearest neighbors for each 
node according to a power law (scale-free property), see,  for 
example, Refs.~\cite{AdHu00,AlJeBa99,FaFaFa99,AlKuSt02}. 
To model these networks and their growth very often 
stochastic models and methods from statistical physics 
have been considered; see Ref.~\cite{AlBa02}. However, the use 
of exact deterministic models allows a quick determination 
of the relevant parameters of the associated graph that may 
be compared with experimental data from real 
and simulated networks.
Previous work, for example, considered
deterministic small-world networks
comparable to those obtained stochastically 
by Watts and Strogatz~\cite{CoOzPe00}, node replacement and node addition
methods to produce small-world and scale-free networks from
a low diameter ``backbone'' network~\cite{CoSa02},
specific recursive scale-free constructions with fixed degree 
distributions~\cite{BaRaVi01,DoGoMe02,RaBa03,No03} and scale-free trees 
(without clustering)~\cite{JuKiKa02}.
In this paper we present a  recursive graph construction
which produces scale-free small-world networks with an
adjustable clustering and such that the parameter 
of the power law
associated with the degree distribution,  $\gamma$,
takes values between $2$ and $1+{\ln 3}/{\ln 2}=2.584 96$.
It has been shown that the values
for the scaling exponent of most real technical networks
~\cite{Me03} are in this range. 
Moreover, the addition of one-degree nodes to our construction
allows a model which is not far from the real
topology of Internet at the autonomous system level, as is
described in Ref.~\cite{TaPaSiFa01}. 

Therefore, we introduce here a deterministic exact network model 
of complex systems for which we can adjust both
the clustering parameter and the power-law exponent, 
providing a complementary tool 
to stochastic approaches.

\begin{figure}
\includegraphics[width=65mm]{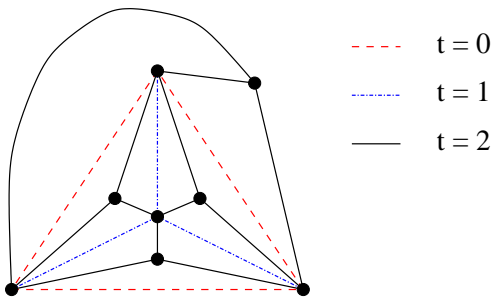}
\caption{\label{fig:q3} (Color online) First stages of a growing recursive clique tree $K(3,t)$. $K(3,0)$ is the complete
graph or clique $K_3$ (the triangle with thick edges). At each step a vertex is connected to each of
existing $K_3$. Therefore at $t=1$ only one vertex is added, but at $t=2$, four new vertices 
should be connected to the four different cliques $K_3$.}
\end{figure}

{\em Recursive clique tree $K(q,t)$, $q\geq 2$}. Definition.---
a complete graph $K_q$ (also referred 
in the literature as $q$ clique; see Ref.~\cite{We01}) is
a graph without loops whose $q$ vertices are pairwise
adjacent.
The recursive clique tree $K(q,t)$ is the graph constructed
as follows:
For $t=0$, $K(q,0)$ is the complete graph $K_q$ (or $q$ clique).
For $t\geq 1$, $K(q,t)$ is obtained from  $K(q,t-1)$ by adding 
for each of its existing subgraphs isomorphic to a $q$-clique a
new vertex and joining it to all the vertices of
this subgraph (see Fig. \ref{fig:q3} for the case $q=3$).
Then, at $t=1$, $K(q,1)$ results in the complete 
graph with $q+1$ vertices, $K_{q+1}$, and at $t=2$ we 
add $q+1$ new vertices 
each of them  connected to all the
vertices of one of the $q$ cliques $K_q$ (subgraphs of $K_{q+1}$), 
and so on.  

This algorithm produces a complex growing 
graph with a tunable parameter $q$
which controls all its relevant characteristics.
In the particular case $q=2$ we obtain the same
graph as in Ref. ~\cite{DoGoMe02}. However, our 
family is infinite as $q$ can take any natural
value starting from 2.

Notice that although we call the graph a recursive clique tree, 
the graph contains numerous cycles 
and hence is not a tree in the strict sense. 
Recursive clique-trees $K(q,t)$ are a natural generalization of trees 
(if one considers the case $q=1$, 
then we obtain a subfamily of  the ``usual'' trees).
Similar constructions in which new vertices 
are joined to every vertex of a given $q$ clique
have been considered, in another context, 
in Ref. ~\cite{Bo88} and termed ``$k$ trees.''

\begin{figure}
\includegraphics[width=80mm]{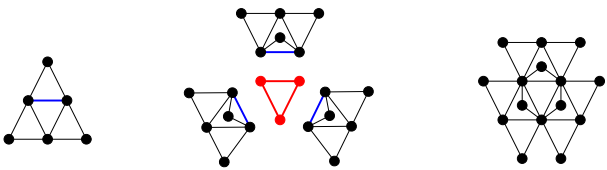}
\caption{\label{fig:recq2}(Color online) Recursive construction of $K(2,t)$.
We {\em glue} a native (or initial) clique of a $K(2, t-1)$ on each 
$2$-clique of $K_3$ to construct $K(2,t)$. In this figure we
obtain $K(2,3)$ from three copies of $K(2,2)$ as follows: The graph at the left is $K(2,2)$ with
its native $2$-clique (the thick edge). Three copies of $K(2,2)$ are glued 
by their natives $2$-cliques
to each of the three $2$-cliques of $K_3$ (center) resulting in $K(2,3)$ (right).}
\end{figure}

{\em Recursive construction of $K(q,t)$}.
There exists another interesting way to construct $K(q,t)$, 
which clearly shows the recursive structure of such networks. 
We call {\em native clique} of  $K(q,t)$ its initial $q$ clique
at $t=0$. Then $K(q,t)$ is constructed as follows:

At step 0, we have the native clique $K_q$.
For any step $t\geq 1$, $K(q,t)$ is constructed as follows: 
Consider a $(q+1)$ clique, then every subgraph of it
isomorphic to a $q$ clique (there are $q+1$ such cliques) 
is a native clique of a $K(q,t-1)$; see Fig.~\ref{fig:recq2}.

{\em Size  and order of $K(q,t)$}.
Table~\ref{tab:table1} gives the number of new edges added
to the tree at each step and the total number of $K_q$
at this step.
\begin{table}
\caption{\label{tab:table1} Number of new edges added to the recursive 
clique tree $K(q,t)$ at each step $t$ and total
number of complete graphs $K_q$ at this step.}
\begin{ruledtabular}
\begin{tabular}{lll}
Step ($t$)  & New edges &  Number of $K_q$\\
\hline
0  &  $\frac{q(q-1)}{2}$ & $1$ \\  
1  &  $q$ & $q+1$ \\   	
2  &  $q(q+1)$ & $q(q+1)+(q+1)=(q+1)^2$ \\  
3  &  $q(q+1)^2$      & $q(q+1)^2+(q+1)^2=(q+1)^3$\\
$\cdots$ & $\cdots$ & $\cdots$ \\
i  &  $q(q+1)^{i-1}$ & $(q+1)^i$ \\
i+1  &  $q(q+1)^{i}$ & $q(q+1)^i+(q+1)^i=(q+1)^{i+1}$ \\
$\cdots$ & $\cdots$ & $\cdots$ \\
\end{tabular}
\end{ruledtabular}
\end{table}
Therefore we can easily compute the total of edges
at step $t$,
\begin{equation}
|E|_t =\frac{q(q-1)}{2} + q\sum_{i=0}^{t-1}(q+1)^i=
\frac{q(q-1)}{2}+(q+1)^{t}-1
\end{equation}
The distribution of vertices and degrees is given in Table~\ref{tab:table2}.
\begin{table}
\caption{\label{tab:table2} Distribution of
vertices and degrees for the recursive 
clique tree $K(q,t)$  at each step $t$.}
\begin{ruledtabular}
\begin{tabular}{lll}
  Step ($t$)  & Num. vertices  &  Degree \\
\hline
1  &  $q+1$ & $q$ \\   	
2  &  $q+1$ & $2q$ \\  
    &  $q+1$ & $q$ \\  
3  &  $q+1$      & $q^2+2q$ \\
   &  $q+1$      & $2q$ \\  
   &  $(q+1)^2$  & $q$ \\  
4 &  $q+1$      & $q^3 +q^2+2q$ \\
  &  $q+1$      & $q^2+2q$ \\
  &  $(q+1)^2$  & $2q$ \\  
  &  $(q+1)^3$  & $q$ \\  
$\cdots$ & $\cdots$ & $\cdots$ \\
i  &  $q+1$ & $q^{i-1} +q^{i-2}+\cdots  + q^2+2q$ \\
   &  $q+1$ & $q^{i-2}+\cdots  + q^2+2q$ \\
   &  $\cdots$  &   \\
   &  $(q+1)^{i-2}$ & $2q$ \\  
   &  $(q+1)^{i-1}$ & $q$ \\ 
$\cdots$ & $\cdots$ & $\cdots$ \\

\end{tabular}
\end{ruledtabular}
\end{table}

The maximum degree at step $i$ is 
\begin{equation}
\Delta_i=\frac{q-q^i}{1-q}+q=\frac{q^i-1}{q-1}+q-1
\end{equation}
Adding up the number of vertices gives the result
\begin{equation}\label{Nt}
N_t=\sum_{j=1}^{t}(q+1)^j+(q+1)=\frac {(q+1 )^{t}-1}{q}+q
\end{equation}
The average degree is then 
\begin{equation}
\overline{k}_t = \frac{2|E|_t}{N_t} = 
 \frac {q [ {q}^{2}-q+2\, \left( q+1 \right) ^{t}-2 ] }
{ \left( q+1 \right) ^{t}-1+{q}^{2}}
\end{equation}
[for $q=2$, it is $4/(1+3^{1-t})$]


{\em Degree distribution}.
The degree spectrum of the graph is discrete: at time $t$, 
the number $N(k,t)$ of vertices of degree 
$k = q, 2q, q^2+2q, q^3+q^2+2q, \ldots, 
q^{t-2}+q^{t-3}+\cdots +q^2+2q, q^{t-1}+q^{t-2}+\cdots +q^2+2q$ 
is equal to 
$(q+1)^{t-1}, (q+1)^{t-2}, (q+1)^{t-3},(q+1)^{t-4}, \ldots, 
(q+1), (q+1)$, 
respectively. 
Other values of the degree are absent. 
Clearly, for the large network, $N(k,t)$ decreases as 
a power of $k$, so the network can be called ``scale free.''
Spaces between degrees of the spectrum grow with increasing $k$. 
Therefore, to relate the exponent of this discrete degree 
distribution to the standard  
$\gamma$ exponent of a continuous degree distribution 
for random scale free networks, 
we use a cumulative distribution 
$P_{cum}(k) \equiv \sum_{k^\prime \geq k} 
N(k^\prime,t)/N_t \sim k^{1-\gamma}$. 

Here $k$ and $k^\prime$ are points of the discrete 
degree spectrum. For a degree 
\begin{equation*}
k = q^{t-l}+q^{t-l-1}+\cdots +q+q
   = q\left(\frac{q^{t-l}-1}{q-1}+1\right)
\end{equation*}
there are  $(q+1)^{l-1}$ vertices
with this exact degree.

We count now  how many vertices have this and a higher degree.
From the distribution
\begin{equation*}
\sum_{k' \geq k} N(k',t)= \sum_{p=1}^{l-1} (q+1)^p + (q+1) =
\frac{(q+1)^l-1}{q} +q
\end{equation*}
As the total number of vertices at step $t$ is given
in Eq.~(\ref{Nt}) we have
\begin{equation*}
\left[q(\frac{q^{t-l}-1}{q-1}+1)\right]^{1-\gamma} =
\frac{\frac{(q+1)^l-1}{q} +q}{ \frac{(q+1)^t-1}{q} +q }
= \frac{(q+1)^l-1+q^2}{(q+1)^t-1+q^2}
\end{equation*}
Therefore, for $t$ large
\begin{equation*}
(q^{t-l})^{1-\gamma}= (q+1)^{l-t}
\end{equation*}
and
\begin{equation}
\gamma \approx 1+\frac{\ln (q+1)}{\ln q}
\end{equation}
so that $2<\gamma <2.584 96$. 

Also, notice that when $t$ gets large, the maximal 
degree of a vertex is roughly 
$q^{t-1} \sim N_t^{\ln q/\ln (q+1)} = N_t^{1/(\gamma-1)}$.
%



{\em Clustering distribution}.
The clustering coefficient $C(x)$ of a vertex $x$
is the ratio of the total number of existing 
connections between all $k$ its nearest neighbors 
and $k(k-1)/2$, the number  of all possible connections between them. 
The clustering of the graph is obtained averaging
over all its vertices. In what follows 
${\cal N}_i(x)$ will denote the total number of neighbors 
which $x$ has at step $i$, ${\cal K}_i(x)$ will denote the number
of different $q$ cliques which contain  $x$ at step $i$, and 
${\cal E}_i(x)$ will denote the total number of edges
among the neighbors of vertex $x$ also at step $i$.

\smallskip

Next we compute the clustering parameter for a vertex $x$ at any
step of iteration.


{\em Step 1.} 
As $x$ belongs to  $K_{q+1}$, i.e., the recursive clique tree $K(q,t)$
at step 1, it is adjacent to $q$ other vertices [$ {\cal N}_1(x) =q$]  which form a $K_q$. 
Therefore the number
of edges among all vertices adjacent to $x$ is
${\cal E}_1(x)={q(q-1)}/{2}$.
Moreover, $x$ belongs to $q$ different $q$ cliques, i.e., 
${\cal K}_1(x)=q$.


{\em Step 2.} 
As the number of $q$ cliques in $K_{q+1}$ is $q+1$,
we add this number of new vertices to the construction.
But of these, $q$ will be forming --each--  a different 
$(q+1)$ clique with $x$. Therefore
${\cal N}_2(x) =2q$ and  ${\cal E}_2(x)={q(q-1)}/{2}+q(q-1)$.
Now $x$ will belong to ${\cal K}_2(x)={\cal K}_1(x)+q(q-1)=q^2$
different $q$ cliques.


{\em Step 3.} 
We add now to the construction  ${\cal K}_2(x)$
new vertices which will form each a $(q+1)$ clique with $x$.
${\cal E}_3(x)={\cal E}_2(x)+ {\cal K}_2(x)(q-1)=
{q(q-1)}/{2}+q(q-1)+q^2(q-1)$.
${\cal K}_3(x)={\cal K}_2(x)+{\cal K}_2(x)(q-1)=q^2+q^2(q-1)=q^3$
and ${\cal N}_3(x)={(q^3-1)}/{(q-1)}+q-1$.


{\em Step t.}
We add ${\cal K}_{t-1}(x)$ new vertices forming each 
a $(q+1)$ clique with $x$.
${\cal E}_{t}(x) ={\cal E}_{t-1}(x) + {\cal K}_{t-1}(x)(q-1)=
{q(q-1)}/{2}+(q-1)\sum_{i=1}^{t-1}q^i=
(q-1)[{q}/{2}+{q^t-1}/{(q-1)}-1]$.
${\cal K}_{t}(x)={\cal K}_{t-1}(x)+{\cal K}_{t-1}(x)(q-1)=
{\cal K}_{t-1}(x)q=q^{t}$
and ${\cal N}_t(x)={(q^t-1)}/{(q-1)}+q-1$.

Therefore the clustering for vertex $x$ after $t$ iterations is
\begin{equation}
C(x)=\frac{2{\cal E}_{t}(x) }{ {\cal N}_t(x) [ {\cal N}_t(x)-1]}
\end{equation}
Notice that $ {\cal N}_t(x) = \Delta_t$ is precisely the maximum degree
at step $t$.

We see easily that there is a one-to-one correspondence
between the degree of a vertex and its clustering.
In general, we will have that for a vertex $v$ of degree $k$
the clustering parameter will be
\begin{equation}
C(v)=\frac{2(q-1)(k-\frac{q}{2})}{k(k-1)}
\end{equation}
Notice that for $q=2$ we have
$C(v)=2/k$ as in Ref. \cite{DoGoMe02}.\\
Using this result, we can compute now the clustering of the graph,

\begin{equation}
\small
\overline{C}_t =\frac{2(q+1)(q-1)(\Delta_t-\frac{q}{2})}{\Delta_t(\Delta_t-1)}
     +  \sum_{i=1}^{t-1} \frac{2(q+1)^{t-i}(q-1)(\Delta_i-\frac{q}{2})}{\Delta_i(\Delta_i-1)}
\end{equation}
For  $t\geq 7 $ and for any $q\geq 3$ we have 
\begin{equation}
\overline{C}_t\geq\frac{3q-2}{3q-1}
\end{equation}

Thus the clustering is high, and, similarly to the $\gamma$ 
coefficient of the power law, is tunable by choosing 
the right value of $q$: in particular, $\overline{C}$ ranges 
from $\frac{4}{5}$ (in the case $q=2$) to a limit of 1 when $q$ gets large.

{\em Diameter}.
We recall that the diameter of a graph is the maximum of all 
distances between any two vertices of the graph.
Computing the exact diameter of $K(q,t)$ can be done analytically, and
gives the result shown below. Here, we present the main ideas of
this analysis.

First, we note that for any step $t$, the diameter always lies between
pair of vertices that have just been created at this step.
We will call such vertices {\em outervertices}. Then, we note
that, by construction, no two outervertices created at the same step
$t$ can be connected. Hence, if $v_t$ is an outervertex created at step $t$,
then $v_t$ has been connected to a $q$ clique composed of vertices
created at pairwise different steps $t_1<t_2<\cdots<t_q$ and consequently, $t_1\leq t-q$.
Now we want to know the distance between two outervertices $u_t$ and
$v_t$. The idea is, starting from $u_t$, to ``get back'' to $K(q,1)$
by jumps from $u_t$ to $u_{t-q}$, then to
$u_{t-2q}$, etc. Hence it takes at most about ${t}/{q}$ jumps to go from
$u_t$ to $K(q,1)$. Then it takes at most as many jumps to go
from $K(q,1)$ to $v_t$ and we conclude that the diameter
cannot be bigger than (roughly) ${2t}/{q}$.

More precisely, the exact formula for the diameter of $K(q,t)$, 
denoted $Diam[K(q,t)]$, is the following~:
\begin{equation}
Diam[K(q,t)]=2\left(\lfloor\frac{t-2}{q}\rfloor +1\right)+f(q,t)
\end{equation}
where $f(q,t)=0$ if $t-\lfloor{(t-2)}/{q}\rfloor q\leq \lceil{(q+1)}/{2}\rceil$, and 1 otherwise.
This value can be obtained by a sharp analytical proof.

When $t$ gets large, then $Diam[K(q,t)]\sim {2t}/{q}$, while
$N_t\sim q^{t-1}$, thus the diameter clearly grows logarithmically
with the number of vertices.

{\em Discussion}.
The recursive clique-trees $K(q,t)$ which we study in this paper are
recursively constructed  graphs which have both
small-world and scale-free characteristics.
By choosing adequately the value of $q$
it is possible to obtain different
clustering parameters and power-law exponents
(the clustering of the graph ranges from $0.8$ to $1$ 
while the power-law exponent
takes values between $2$ and $2.584 96$).
It has been shown  that many networks modeling complex systems such as
the WWW, Internet, movie actors, the Erd\H{o}s number, interactions in protein complexes
of {\em Saccharomyces cerevisae}, the metabolic network of {\em Escherichia coli}, etc.,
have clustering and power-law parameters precisely in these ranges, see, for 
example, 
Refs. \cite{BaDeRaYoOl03,FaFaFa99,WaSt98,Me03,JeToAlOlBa00,GoOhJeKaKi02}. 
Moreover, the recursive clique trees are actually a 
deterministic tunable generalization of the 
scale-free growing networks introduced in Ref.~\cite{DoMeSa01} in which 
one vertex is created per unit time and connects 
to both the ends of a randomly chosen edge. These networks were further
generalized and called ``pseudofractal'' graph
in \cite{DoGoMe02}, and  correspond to
the particular case $q=2$ of the recursive clique trees. 
Variations of our construction, for example, by choosing
that in different steps the new vertices added are attached
to cliques of different size, would allow a
richer structure and  more flexibility in the 
control of the clustering and power-law exponent and other relevant 
network parameters.
\smallskip

The authors thank a Picasso bilateral grant, references
HF2001-0056 (Spain) and 04295SG (France). 
Additional support for the first author was
provided by the Ministry of Science and Technology, Spain, and the
European Regional Development Fund (ERDF) under Project No.
TIC2002-00155.

\end{document}